\documentclass[sn-mathphys,Numbered]{sn-jnl}

\usepackage{graphicx}%
\usepackage{multirow}%
\usepackage{amsmath,amssymb,amsfonts}%
\usepackage{amsthm}%
\usepackage{mathrsfs}%
\usepackage[title]{appendix}%
\usepackage{xcolor}%
\usepackage{textcomp}%
\usepackage{manyfoot}%
\usepackage{booktabs}%
\usepackage{algorithm}%
\usepackage{algorithmicx}%
\usepackage{algpseudocode}%
\usepackage{listings}%
\usepackage{upgreek}
\providecommand{\dd}{\mathop{}\!\mathrm{d}}

\begin{document}

\title{Bessel--Gaussian Beam Propagation in a Thermally Induced Axially Varying GRIN Medium}


\author{\fnm{Fatemeh} \sur{Kalantarifard}}

\author{\fnm{Parviz} \sur{Elahi}}

\email{parviz.elahi@ozyegin.edu.tr}

\affil{Department of Natural and Mathematical Sciences, Özyeğin University, 34794 Istanbul, Türkiye}


\abstract{High-power end-pumped solid-state lasers often operate in regimes where pump-induced heating creates a strong refractive-index gradient (thermal lensing) that governs resonator stability and mode quality. When the pump is absorbed according to the Beer--Lambert law, the thermal load, and hence the GRIN strength, vary along the crystal length, so the standard ABCD matrix of a constant-gradient GRIN element is no longer directly applicable. Here we derive a closed-form ABCD transmission matrix for a thermally loaded laser crystal pumped by a top-hat beam while explicitly accounting for axial absorption. Starting from the steady-state heat equation, we obtain the temperature field and the associated thermo-optic index profile. We then solve the paraxial eikonal ray equation analytically and express the transfer-matrix elements in terms of Bessel and Neumann functions. The resulting matrix is validated against the conventional slab-product method and shown to recover the uniform-medium and constant-gradient GRIN limits. Finally, we illustrate its utility by modeling Bessel--Gaussian beam propagation through the axially varying thermally induced GRIN medium.}




\maketitle
\section{Introduction}
Thermal loading is one of the dominant factors limiting power scaling in pumped lasers. As the absorbed pump power increases, heat generation in the gain medium produces temperature gradients that degrade optical properties through thermal lensing, thermally induced stress, and stress-induced birefringence \cite{Koechner2006, clarkson2001thermal,huang2007thermo, brown2002thermal}. These thermo-optic and thermo-mechanical effects ultimately compromise resonator stability, beam quality, and efficiency, and they become particularly severe in high-power end-pumped architectures, where heat deposition is strongly localized.

The magnitude and spatial distribution of thermal effects depend on several coupled parameters, including the pump-beam intensity profile, the pumping configuration (e.g., single-end vs double-end pumping), and the geometry and boundary conditions of the active medium (rod, slab, thin-disk, composite structures, and cooling scheme). Regarding the pump profile, non-ideal diode beams and fiber-coupled delivery often lead to deviations from a simple Gaussian distribution, which in turn modify the heat-source term and the resulting temperature and stress fields. In this context, H. Nadgaran et al. \cite{nadgaran2005analytical} investigated how non-uniform heat deposition influences the temperature distribution, mechanical stress, and thermo-optic behavior, including the temperature-dependent refractive-index change and thermally induced stress contributions to optical path distortion. Building on the need for realistic pump descriptions, Y. Wang et al. \cite{wang2023combined} introduced a combined super-Gaussian pump model and reported improved agreement with measured pump profiles by fitting experimental beam-intensity data from a fiber-coupled laser diode, thereby enhancing the predictive accuracy of thermal-lens estimations. Similarly, P. Shang et al. \cite{shang2023numerical} performed finite-element numerical simulations of thermal effects in laser-diode end-pumped solid-state lasers using a super-Gaussian pump profile, demonstrating that the choice of pump shape alters peak temperature, radial gradients, and stress distributions, and thus the severity of thermal aberrations.

The pumping scheme also plays a critical role by altering the axial symmetry and the peak thermal load. A substantial body of work has examined how double-end pumping can reduce axial temperature asymmetry, lower the maximum temperature rise, and mitigate thermal lensing compared with single-end pumping, especially when absorption is strong, and the heat load is concentrated near the entrance face \cite{elahi2010double,fang2015thermal,liu2016analytical,huang2014analysis,li2010thermal,vorobyev2025optimization,kalantarifard2009analytical}. The thermal response is also strongly affected by the active-medium geometry and its heat-removal pathways. For example, slab geometries can promote more efficient heat extraction and provide a degree of control over thermal gradients through quasi-one-dimensional heat flow, while also introducing distinct stress patterns and potential polarization issues that must be managed through optical design and crystal orientation \cite{tilleman2011analysis,lyashedko2011thermo,elahi2011calculation,cheng2019research}. In contrast, rod-like media tend to develop stronger radial thermal gradients under end pumping, whereas thin-disk concepts distribute the heat load over a large cooled surface to suppress thermal lensing. 

Under end pumping, the nonuniform temperature distribution generated inside the gain medium induces a spatial variation of the refractive index through both thermo-optic and stress-related contributions. In the paraxial region, this thermally induced index profile is commonly approximated by a quadratic radial dependence, so the heated crystal can be treated as an effective gradient-index (GRIN) element rather than as an ideal thin lens \cite{Koechner2006}. 
Beyond its role in modeling thermal lensing, the GRIN medium has also been widely used as a canonical platform for studying the propagation dynamics of structured and partially coherent beams. Previous investigations have examined Gaussian vortex beams \cite{Yang2020GaussianVortexGRIN}, Airy--Gaussian vortex beams \cite{Zhao2016AiryGaussianVortexGRIN}, Lorentz--Gauss vortex beams \cite{Qusailah2023LorentzGaussVortexGRIN}, standard and elegant Laguerre--Gaussian vortex beams \cite{Odzak2025LaguerreGaussianGRIN}, partially coherent generalized Hermite cosh--Gaussian beams \cite{Saad2024PCGHCGaussianGRIN}, Bessel--Gaussian beams \cite{Pei2019BesselGaussianGRIN}, partially coherent modified Bessel--Gauss beams \cite{Dong2020ModifiedBesselGaussGRIN}, and both on-axis and off-axis Bessel beams \cite{Cao2018OnOffAxisBesselGRIN} in GRIN media. Collectively, these studies show that the parabolic index profile of a GRIN medium gives rise to characteristic periodic focusing and defocusing behavior, beam reconstruction, and phase evolution, while the detailed propagation features remain strongly dependent on beam order, topological charge, coherence, waist size, and launch conditions such as decentering and tilt. This broader body of work highlights the importance of compact analytical propagation models for GRIN systems, especially when one seeks to connect thermally induced index gradients in laser crystals with the evolution of nontrivial beam families.
 
In standard analyses of GRIN media, the refractive-index distribution is usually assumed to be independent of the longitudinal coordinate $z$, so the gradient parameter remains constant along the propagation direction. This approximation permits the use of the conventional ABCD matrix for a uniform GRIN medium. By contrast, when the refractive-index perturbation varies axially, as in end-pumped laser crystals with longitudinally decaying heat deposition, the GRIN strength becomes an explicit function of $z$. In this more general case, propagation is commonly modeled using the slab-product method, in which the medium is discretized into many thin slices, each approximated as a uniform GRIN segment, and the total transfer matrix is constructed from the product of the individual slice matrices \cite{sabaeian2008bessel}.

In this work, we derive a compact, closed-form transmission matrix for an end-pumped laser crystal with a top-hat pump profile and Beer--Lambert absorption. We first obtain the steady-state temperature field and the corresponding thermo-optic index profile, then solve the paraxial eikonal (ray) equation analytically to obtain the ABCD elements for a GRIN medium whose strength decays exponentially along $z$. We validate the resulting matrix by comparison with the slab-product method and by verifying the symplectic conditions. As an application, we model the propagation of Bessel--Gaussian beams through the axially varying thermally induced GRIN medium.

\section{Temperature distribution and refractive index variation}
The steady-state heat conduction equation with temperature-independent thermal conductivity $k$ is
\begin{equation}
k\nabla^2 T(r,z)+S(r,z)=0,
\end{equation}
where $S(r,z)$ is the thermal power density associated with heat dissipated in the gain medium. For an end-pumped configuration with a top-hat pump profile, the volumetric heat source is
\begin{equation}
S(r,z)=
\begin{cases}
\dfrac{\eta P_{in}\alpha e^{-\alpha z}}{\pi a^2\left(1-e^{-\alpha L}\right)} & 0\le r\le a,\\[6pt]
0 & a<r\le b,
\end{cases}
\end{equation}
where $\eta$ is the fractional thermal load, $P_{in}$ is the incident pump power, $\alpha$ is the absorption coefficient, $L$ is the crystal length, $a$ is the pump radius, and $b$ is the crystal radius. We assume forced edge cooling by a coolant held at a fixed temperature $T_c$ at the outer boundary $r=b$. Under strong edge cooling and for $\alpha a\ll1$, the axial heat flux is negligible compared with the radial heat flux; therefore, the axial derivative term can be neglected relative to the radial derivatives.

The temperature distributions inside and outside the pumped region, denoted by $T_1(r,z)$ and $T_2(r,z)$, can then be written as
\begin{equation}
T_1(r,z)=T_c+\frac{S_{0}a^2}{4k}\left[1+2\ln\!\left(\frac{b}{a}\right)+\frac{2k}{bh}\right]e^{-\alpha z}-\frac{S_{0}}{4k} r^2 e^{-\alpha z},\qquad 0\le r\le a,
\end{equation}
\begin{equation}
T_2(r,z)=T_c+\frac{S_{0}a^2}{2k}\left[\ln\!\left(\frac{b}{a}\right)+\frac{k}{bh}\right]e^{-\alpha z}-\frac{S_{0}a^2}{2k}\ln\!\left(\frac{r}{a}\right)e^{-\alpha z},\qquad a<r\le b,
\end{equation}
where $h$ is the heat transfer coefficient describing heat transfer between the coolant and the crystal edge, and
\begin{equation}
S_{0}=\frac{\eta \alpha P_{in}}{\pi a^2\left(1-e^{-\alpha L}\right)}.
\label{eq:S0_def}
\end{equation}

The temperature-induced refractive-index change in the pumped region can be written as
\begin{equation}
\Delta n_T(r,z)=\big(T_1(r,z)-T_0(z)\big)\frac{\dd n}{\dd T},
\end{equation}
where $\Delta n_T=n-n_0$ and $T_0(z)=T_1(r=0,z)$ is the on-axis temperature. Near the optical axis, the refractive index therefore assumes the standard parabolic GRIN form with an axially varying curvature.

\section{Ray tracing and closed-form transmission matrix}
\label{sec:ABCD_closedform}
The trajectory of a ray in an inhomogeneous medium follows from Fermat's principle and can be written in eikonal form as
\begin{equation}
\frac{\dd}{\dd s}\left[n(\mathbf{r})\frac{\dd\mathbf{r}}{\dd s}\right]=\nabla n(\mathbf{r}),
\end{equation}
where $\mathbf{r}$ denotes the ray position, $s$ is the arc length along the ray, and $n(\mathbf{r})$ is the refractive index. Under the paraxial approximation for rays propagating predominantly along the $z$ direction, and for slowly varying axial dependence ($\alpha a\ll1$), the radial ray equation becomes
\begin{equation}
\frac{\dd^2 r}{\dd z^2}\approx \frac{1}{n}\frac{\partial n}{\partial r}\approx \frac{1}{n_0}\frac{\partial n}{\partial r}.
\end{equation}
Using the thermo-optic index variation in the pumped region, the ray equation reduces to
\begin{equation}
\frac{\dd^2 r}{\dd z^2}+g\,e^{-\alpha z}\,r=0,
\end{equation}
where
\begin{equation}
g=\frac{S_{0}}{2k n_0}\frac{\dd n}{\dd T}.
\label{eq:grin_strength}
\end{equation}

Introducing the change of variables $u(z)=e^{-\alpha z/2}$ and treating $r$ as a function of $u$, we obtain \cite{nadgaran2008calculation}
\begin{equation}
u^2\frac{\dd^2 r}{\dd u^2}+u\frac{\dd r}{\dd u}+\left(\frac{4g}{\alpha^2}\right)u^2 r=0,
\end{equation}
which is the Bessel differential equation of order zero. Its general solution is
\begin{equation}
r(u)=A_0\,J_0\!\left(\frac{2\sqrt g}{\alpha}u\right)+B_0\,Y_0\!\left(\frac{2\sqrt g}{\alpha}u\right),
\end{equation}
where $J_0$ and $Y_0$ are Bessel functions of the first and second kind of order zero, respectively, and $A_0$ and $B_0$ are constants. Let $z=0$ denote the input plane, where the position $r_0$ and slope $r'_0$ are known. Applying these boundary conditions and evaluating the solution at an arbitrary position $z$, we obtain the ABCD elements of the transmission matrix:
\begin{equation}
A=\frac{\sqrt g\,\pi}{\alpha}\left[J_1\!\left(\frac{2 \sqrt g}{\alpha}\right)Y_0\!\left(\frac{2\sqrt g}{\alpha}e^{-\alpha z/2}\right)-Y_1\!\left(\frac{2\sqrt g}{\alpha}\right)J_0\!\left(\frac{2\sqrt g}{\alpha}e^{-\alpha z/2}\right)\right],
\label{eq:A_closedform}
\end{equation}
\begin{equation}
B=\frac{\pi}{\alpha}\left[Y_0\!\left(\frac{2 \sqrt g}{\alpha}\right)J_0\!\left(\frac{2\sqrt g}{\alpha}e^{-\alpha z/2}\right)-J_0\!\left(\frac{2\sqrt g}{\alpha}\right)Y_0\!\left(\frac{2\sqrt g}{\alpha}e^{-\alpha z/2}\right)\right],
\label{eq:B_closedform}
\end{equation}
\begin{equation}
C=\frac{\pi g}{\alpha}e^{-\alpha z/2}\left[J_1\!\left(\frac{2 \sqrt g}{\alpha}\right)Y_1\!\left(\frac{2\sqrt g}{\alpha}e^{-\alpha z/2}\right)-Y_1\!\left(\frac{2\sqrt g}{\alpha}\right)J_1\!\left(\frac{2\sqrt g}{\alpha}e^{-\alpha z/2}\right)\right],
\label{eq:C_closedform}
\end{equation}
\begin{equation}
D=\frac{\sqrt g\,\pi}{\alpha}e^{-\alpha z/2}\left[Y_0\!\left(\frac{2\sqrt g}{\alpha}\right)J_1\!\left(\frac{2 \sqrt g}{\alpha} e^{-\alpha z/2}\right)-J_0\!\left(\frac{2\sqrt g}{\alpha}\right) Y_1\!\left(\frac{2 \sqrt g}{\alpha} e^{-\alpha z/2}\right)\right].
\label{eq:D_closedform}
\end{equation}
Equations~\eqref{eq:A_closedform}--\eqref{eq:D_closedform} provide the transmission matrix for a GRIN medium whose gradient coefficient varies exponentially along the propagation axis.

\section{Validation and asymptotic behavior}
To validate the derived transmission matrix for an axially varying GRIN medium, we perform two independent checks.

\paragraph{(a) Comparison with the slab-product method.}
We compute a reference matrix by dividing the crystal into $N$ thin slabs. Within each slab, the GRIN coefficient is assumed constant and the corresponding constant-$\gamma$ GRIN matrix is used:
\begin{equation}
M_n=
\begin{pmatrix}
\cos(\gamma_n \Delta z) & \dfrac{1}{\gamma_n}\sin(\gamma_n \Delta z)\\[6pt]
-\gamma_n \sin(\gamma_n \Delta z) & \cos(\gamma_n \Delta z)
\end{pmatrix},
\end{equation}
where $\Delta z=L/N$ and $\gamma_n=\sqrt{g}\,e^{-\alpha z_n/2}$ is evaluated at representative positions $z_n=(n-\tfrac{1}{2})\Delta z$. The overall slab-product matrix is then
\begin{equation}
M=\prod_{n=1}^{N} M_n.
\end{equation}
As $N$ increases, the slab-product elements converge to the closed-form values given by Eqs.~\eqref{eq:A_closedform}--\eqref{eq:D_closedform}.

\begin{figure}[t]
    \centering
    \includegraphics[width=\linewidth]{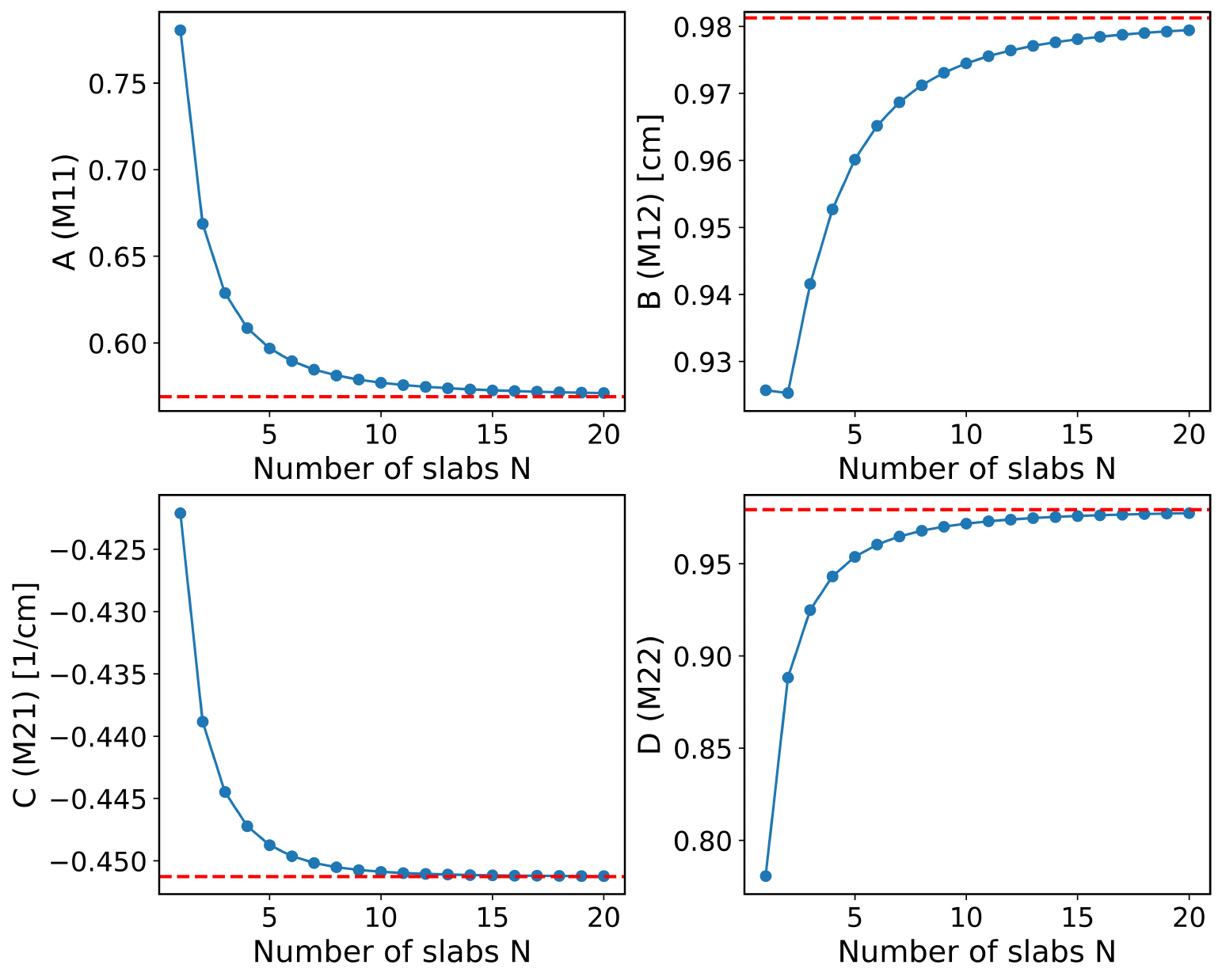}
    \caption{Convergence of the slab-product ABCD matrix (slab slicing) to the closed-form solution in Eqs.~\eqref{eq:A_closedform}--\eqref{eq:D_closedform} for a $L=1$~cm crystal with axially varying GRIN strength due to absorption. Symbols correspond to the slab method with $N$ uniform segments, while the red dashed lines show the closed-form values of $A$, $B$, $C$, and $D$ evaluated at $z=L$.}
    
    \label{fig:Comparison}
\end{figure}

Figure~\ref{fig:Comparison} demonstrates the convergence of the slab-product approximation to the closed-form ABCD matrix as the number of slices $N$ increases. For small $N$, the slab method produces noticeable deviations in all four matrix elements. As $N$ becomes larger, the discretized model more accurately captures the continuous longitudinal change of the refractive-index gradient, and the computed values of $A$, $B$, $C$, and $D$ approach the analytical results shown by the red dashed lines.

\begin{figure}[t]
    \centering
    \includegraphics[width=\linewidth]{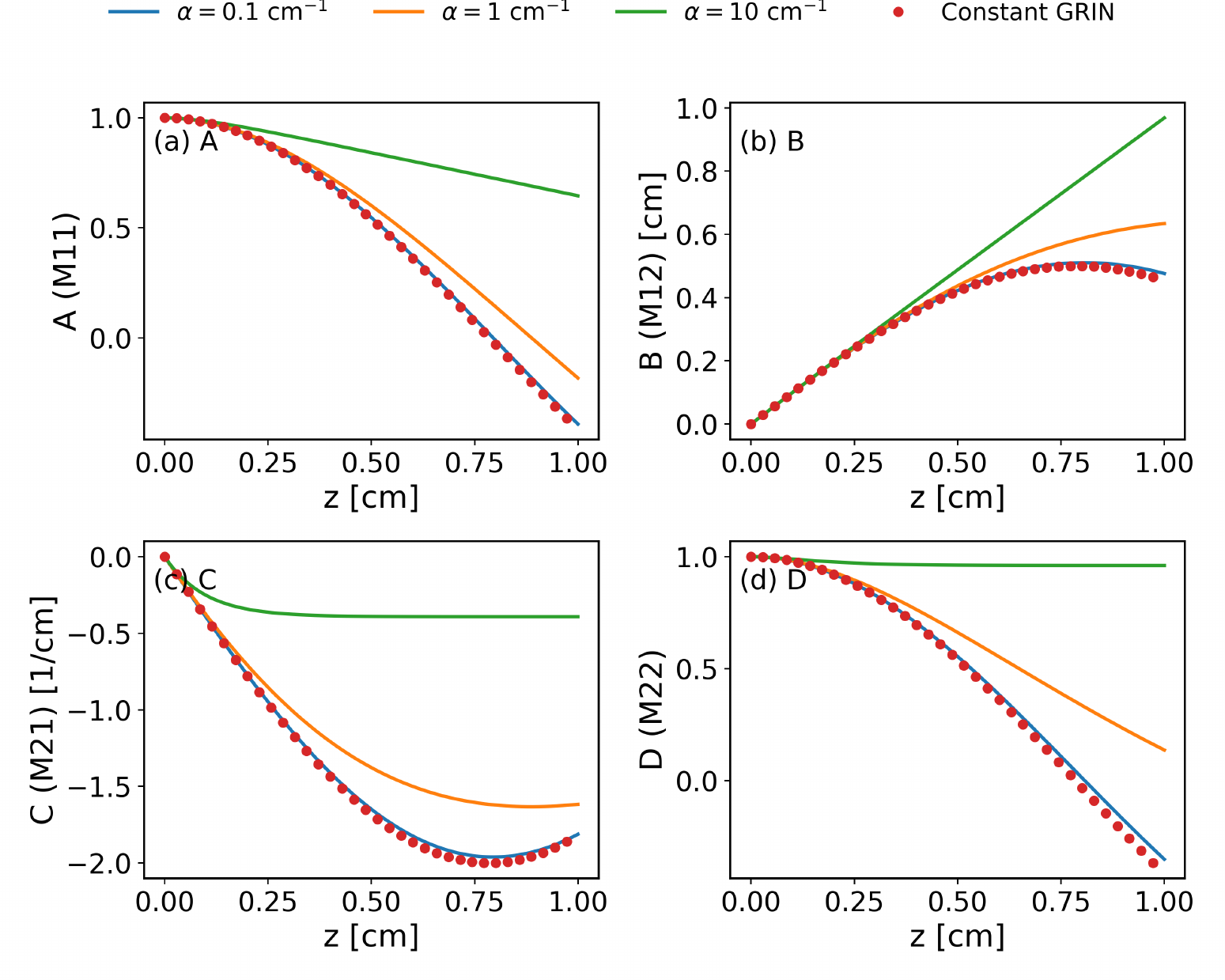}
    \caption{Closed-form ABCD elements of an exponential-decay GRIN medium compared with the standard constant-GRIN model. Panels (a)--(d) show the longitudinal evolution of the matrix elements $A(z)$, $B(z)$, $C(z)$, and $D(z)$ obtained from Eqs.~\eqref{eq:A_closedform}--\eqref{eq:D_closedform} for a fixed GRIN parameter $g=4~\mathrm{cm^{-2}}$ (i.e., $\gamma=\sqrt{g}=2~\mathrm{cm^{-1}}$). Solid curves correspond to absorption coefficients $\alpha=0.1$, $1$, and $10~\mathrm{cm^{-1}}$, while circle markers denote the standard ABCD elements of a constant GRIN medium with the same $\gamma$.}

\label{fig:ABCD_alpha_sweep_g4}
\end{figure}

\paragraph{(b) Limiting cases and consistency checks.}
In the low-pump-power limit $P_{in}\to 0$ (hence $g\to 0$), the thermal index perturbation vanishes and the medium approaches a uniform-index element: 
\begin{equation}
\begin{pmatrix}
A & B\\ C & D
\end{pmatrix}
\;\xrightarrow[P_{in}\to 0]{}\;
\begin{pmatrix}
1 & z\\ 0 & 1
\end{pmatrix}.
\end{equation}

Conversely, for negligible axial variation ($\alpha z\ll 1$), the exponential factor varies weakly and the matrix reduces to the standard constant-gradient GRIN form with $\gamma=\sqrt{g}$:

\begin{equation}
\begin{pmatrix}
A & B\\ C & D
\end{pmatrix}
=
\begin{pmatrix}
\cos(\gamma z) & \dfrac{1}{\gamma}\sin(\gamma z)\\[6pt]
-\gamma\sin(\gamma z) & \cos(\gamma z)
\end{pmatrix}
+\mathcal{O}(\alpha z).
\end{equation}

A detailed asymptotic analysis for the limits $P_{in}\to 0$ and $\alpha z\ll1$ is provided in Appendix~\ref{app:asymptotic}.

Figure~\ref{fig:ABCD_alpha_sweep_g4} shows the variation of the ABCD matrix elements with $z$ for different absorption coefficients and compares the results with those of a conventional GRIN rod. For the smallest absorption coefficient ($\alpha=0.1~\mathrm{cm^{-1}}$), the exponential factor $e^{-\alpha z}$ varies slowly across the 1~cm crystal, so the closed-form elements closely follow the constant-GRIN prediction (circle markers), consistent with the $\alpha z\ll 1$ limit. As $\alpha$ increases to $1$ and $10~\mathrm{cm^{-1}}$, the GRIN strength decays more rapidly with $z$, and the system evolves from strongly focusing near the input face toward a weaker-gradient, nearly uniform medium downstream. This axial weakening reduces the effective phase advance of the GRIN dynamics, which manifests as clear departures from the trigonometric behavior of the constant-GRIN case in all four elements: $A$ and $D$ deviate from $\cos(\gamma z)$, $B$ deviates from $\sin(\gamma z)/\gamma$, and $C$ becomes less negative than $-\gamma\sin(\gamma z)$.
\section{Propagation of a Bessel--Gaussian beam in an exponentially decaying GRIN medium}
\label{sec:BG_exponential_GRIN}
As an application of the closed-form transmission matrix, we consider the propagation of a zero-order Bessel--Gaussian beam in an exponentially decaying GRIN medium.
At the entrance plane $z=0$, we take the cylindrically symmetric zero-order BG beam to be
\begin{equation}
U(r,0)=U_0\,J_0(k_\perp r)\exp\!\left(-\frac{r^2}{w_0^2}\right),
\label{eq:bg_input}
\end{equation}
where $J_0(\cdot)$ is the zeroth-order Bessel function, $w_0$ is the Gaussian envelope radius, and $k_\perp=k\sin\theta$ is the transverse wave number (cone angle $\theta$). Here $k=2\pi n_0/\lambda$ denotes the wavenumber inside the crystal.

\begin{figure}[t]
    \centering
    \includegraphics[width=\linewidth]{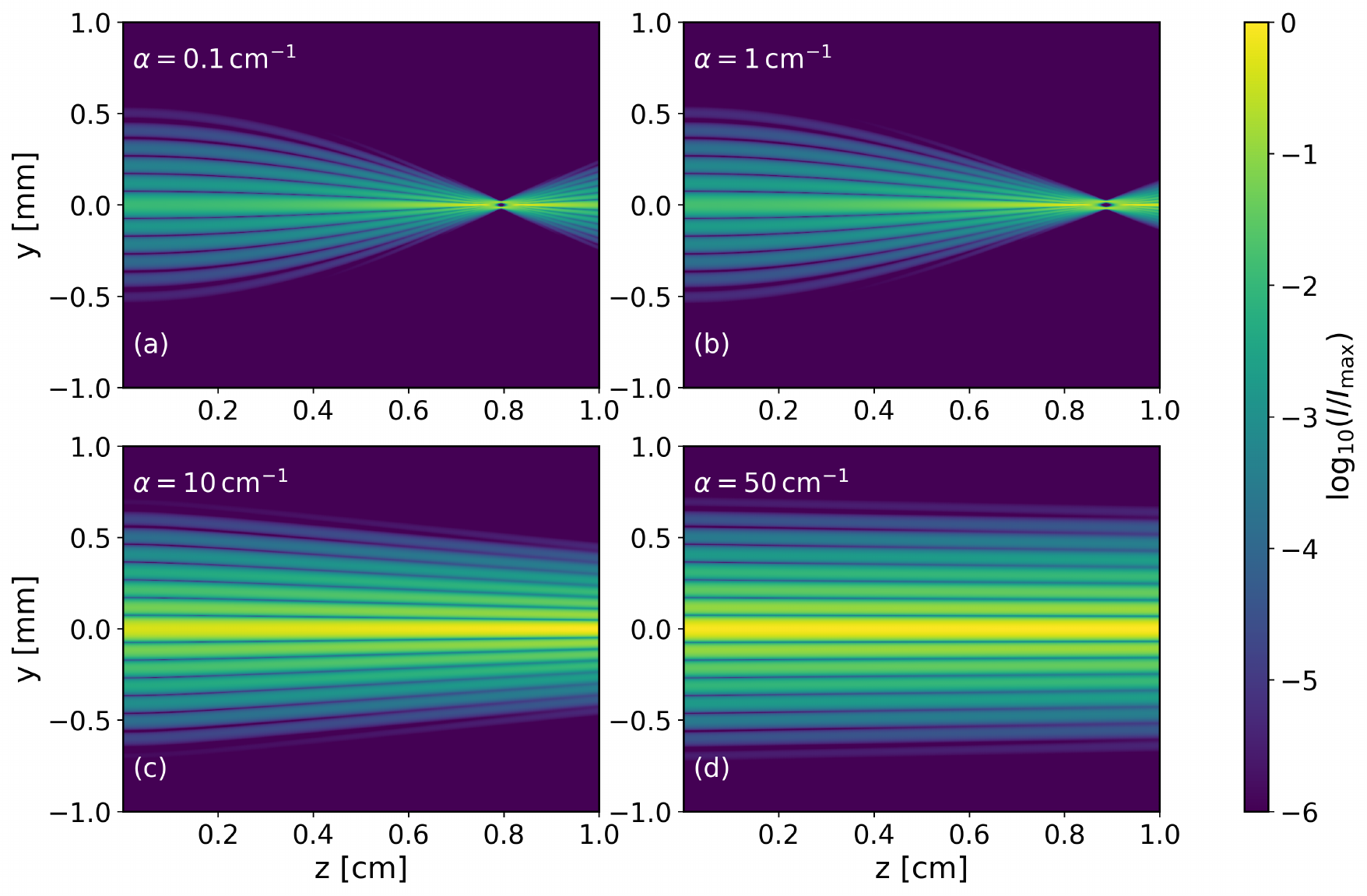}
    \caption{Propagation of a zero-order Bessel--Gaussian beam in an exponential-decay GRIN medium for different absorption coefficients, $\alpha$.
    The intensity distribution in the $y$--$z$ plane is calculated from \eqref{eq:bg_field_closedform} using the closed-form ABCD elements.
    All panels use the same GRIN parameter and crystal length, $g=4~\mathrm{cm^{-2}}$ and $L=1~\mathrm{cm}$, with an input Gaussian envelope radius
    $\omega_0=330~\mu\mathrm{m}$ and cone angle $\theta=0.003$~rad ($m=0$).
    The color scale shows $\log_{10}(I/I_{\max})$, where $I_{\max}$ is the maximum intensity within each panel.}
    \label{fig:Fig3}
\end{figure}
\begin{figure}[t]
    \centering
    \includegraphics[width=\linewidth]{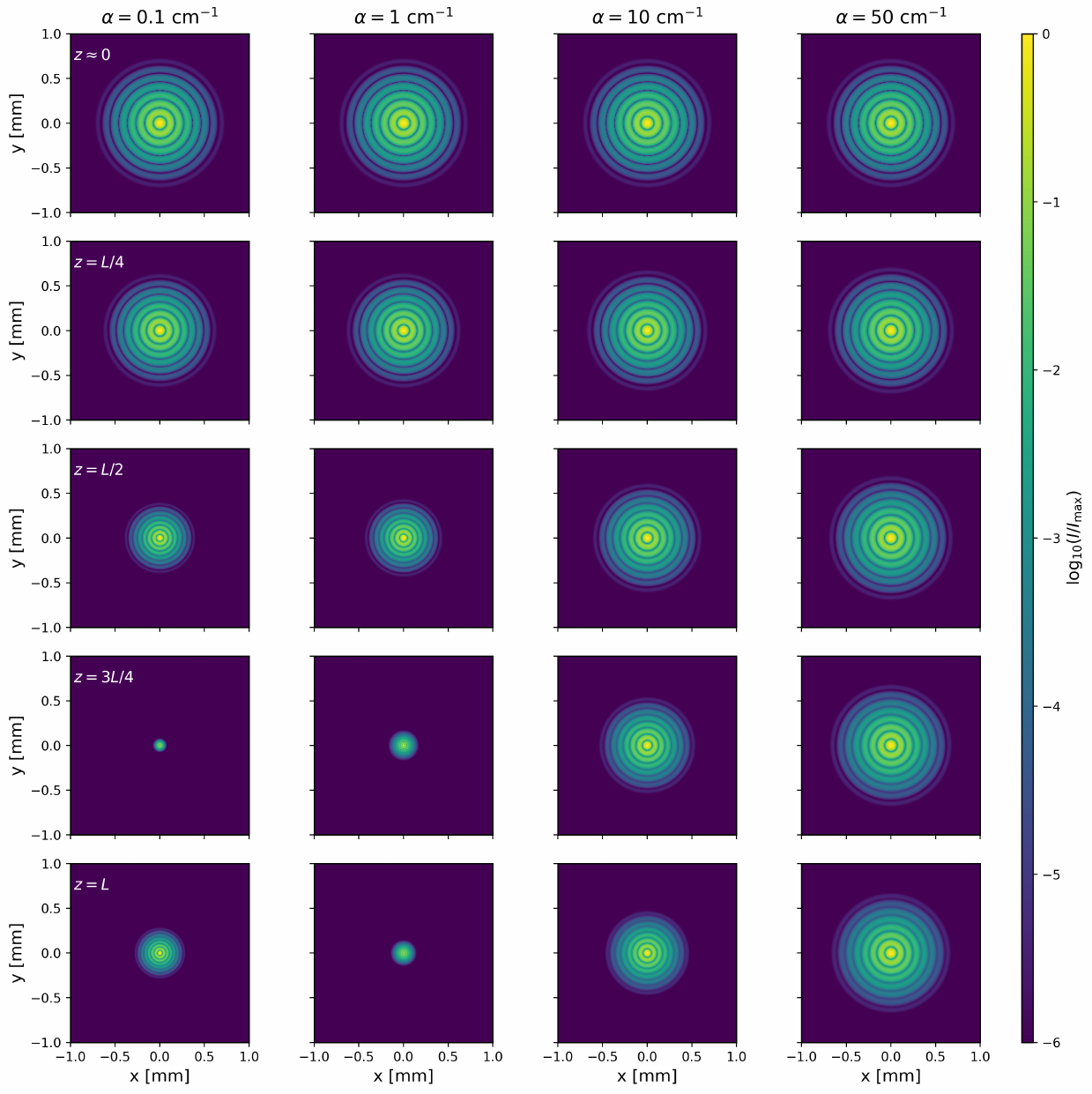}
    \caption{Transverse intensity patterns of the zero-order Bessel--Gaussian beam in an exponential-decay GRIN medium at four representative propagation planes for different absorption coefficients. All panels use the same GRIN parameter and crystal length, $g=4~\mathrm{cm^{-2}}$ and $L=1~\mathrm{cm}$, with an input Gaussian envelope radius $\omega_0=330~\mu\mathrm{m}$ and cone angle $\theta=0.003$~rad ($m=0$).}
    \label{fig:Fig5}
\end{figure}

For an axially symmetric input, the Collins diffraction integral reduces to the cylindrical form derived in Appendix~\ref{app:collins_cyl}:
\begin{equation}
U(r,z)=\frac{k}{iB(z)}
\exp\!\left(i\frac{kD(z)}{2B(z)}\,r^2\right)
\int_{0}^{\infty}
U(\rho,0)\,
\exp\!\left(i\frac{kA(z)}{2B(z)}\,\rho^2\right)
J_0\!\left(\frac{k r\rho}{B(z)}\right)\rho\,\dd\rho.
\label{eq:collins_cyl_main}
\end{equation}

\subsection{Closed-form BG field}
Substituting the BG input from \eqref{eq:bg_input} into the Collins integral in \eqref{eq:collins_cyl_main} yields an integral containing a product of two zeroth-order Bessel functions and a Gaussian factor. Using the identity
\begin{equation}
\int_0^\infty \rho\,e^{-a\rho^2}\,J_0(b\rho)\,J_0(c\rho)\,\dd\rho
=\frac{1}{2a}\exp\!\left(-\frac{b^2+c^2}{4a}\right)I_0\!\left(\frac{bc}{2a}\right),
\qquad \Re(a)>0,
\end{equation}
where $I_0(\cdot)$ is the modified Bessel function of the first kind. The BG field at an output plane located at distance $z$ becomes
\begin{equation}
U(r,z)=
\frac{k}{iB(z)}\,
\frac{1}{2\,a(z)}\,
\exp\!\left(i\frac{kD(z)}{2B(z)}\,r^2\right)\,
\exp\!\left[-\frac{k_\perp^2+c(z)^2}{4\,a(z)}\right]\,
I_0\!\left(\frac{k_\perp\,c(z)}{2\,a(z)}\right),
\label{eq:bg_field_closedform}
\end{equation}
with
\begin{equation}
a(z)=\frac{1}{w_0^2}-i\frac{kA(z)}{2B(z)},
\qquad
c(z)=\frac{k r}{B(z)}.
\end{equation}
This expression enables direct evaluation of the BG intensity distribution $I(r,z)=|U(r,z)|^2$ using the closed-form matrix elements in Eqs.~\eqref{eq:A_closedform}--\eqref{eq:D_closedform}, without axial slicing.

Figure~\ref{fig:Fig3} visualizes how axial absorption reshapes Bessel--Gaussian beam evolution in a thermally induced GRIN medium whose strength decays as $e^{-\alpha z}$. For the smallest absorption coefficient [Fig.~\ref{fig:Fig3}(a), $\alpha=0.1~\mathrm{cm^{-1}}$], the GRIN parameter varies slowly over the 1~cm crystal and the beam exhibits a pronounced, quasi-periodic refocusing/self-imaging behavior typical of a GRIN element. As $\alpha$ increases to $1$ and $10~\mathrm{cm^{-1}}$ [Fig.~\ref{fig:Fig3}(b,c)], the refractive-index curvature decreases more rapidly with propagation, reducing the accumulated GRIN phase and shifting the axial positions and contrast of the intensity revivals; the beam progressively transitions from strong focusing near the input to weaker guiding downstream. In the extreme case $\alpha=50~\mathrm{cm^{-1}}$ [Fig.~\ref{fig:Fig3}(d)], the GRIN action is confined to a short entrance region and the subsequent evolution is dominated by near-free-space propagation, leading to a markedly weakened longitudinal modulation.

Figure~\ref{fig:Fig5} complements the longitudinal maps by showing representative transverse intensity patterns at selected propagation planes. As the absorption coefficient increases, the beam profiles broaden more rapidly downstream and the contrast of the outer rings decreases, reflecting the progressive confinement of the GRIN action to the input region.

\begin{figure}[t]
    \centering
    \includegraphics[width=\linewidth]{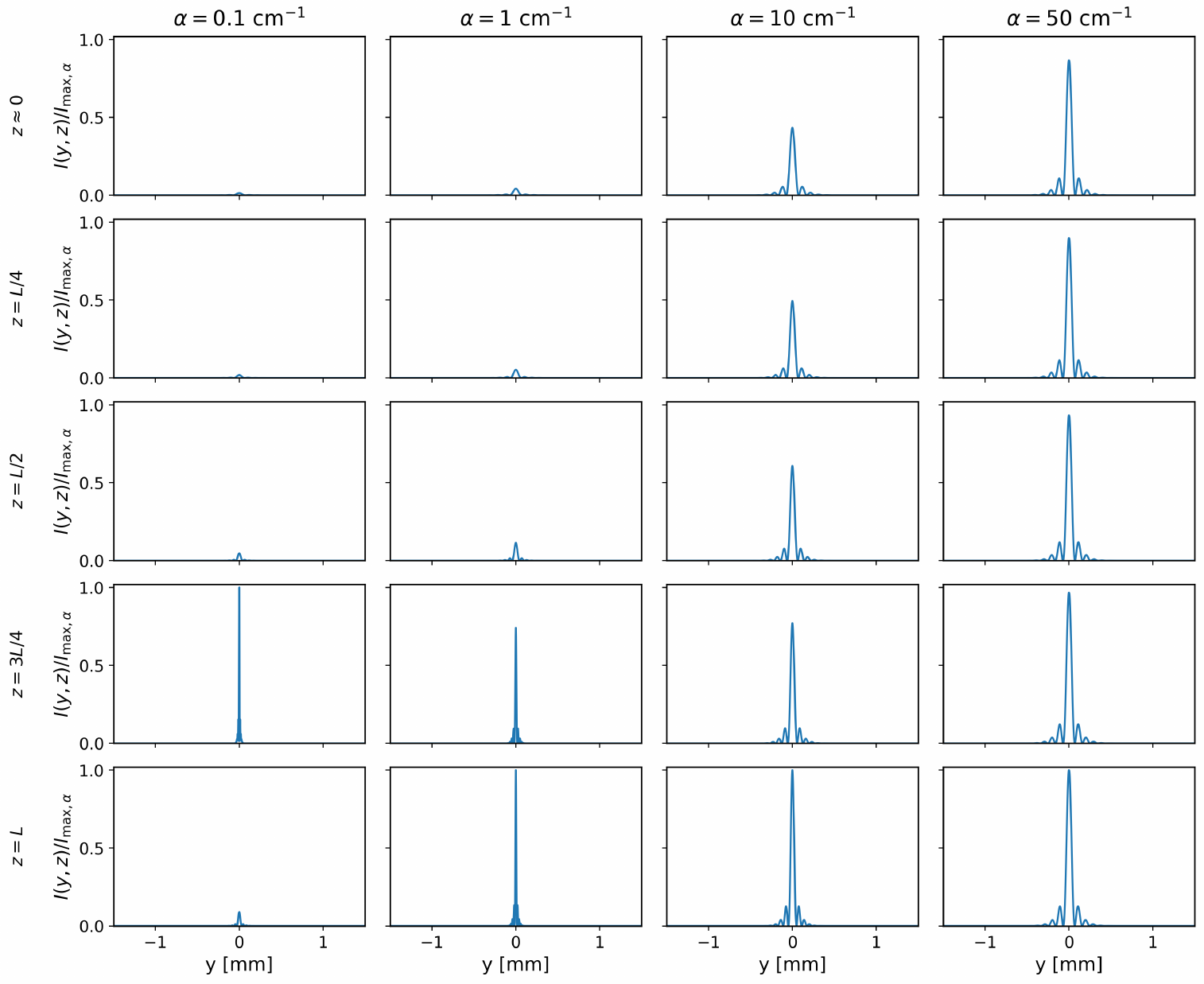}
    \caption{Representative transverse intensity distributions of the zero-order Bessel--Gaussian beam in an exponential-decay GRIN medium at four propagation planes for different absorption coefficients. All panels use $g=4~\mathrm{cm^{-2}}$ and $L=1~\mathrm{cm}$, with $\omega_0=330~\mu\mathrm{m}$ and cone angle $\theta=0.003$~rad ($m=0$).}
    \label{fig:Fig4}
\end{figure}

Figure~\ref{fig:Fig4} further illustrates how the transverse structure of the Bessel--Gaussian beam evolves when the GRIN strength decreases along the propagation axis due to absorption. For weak absorption ($\alpha=0.1~\mathrm{cm^{-1}}$), the refractive-index curvature changes slowly over $L=1$~cm, and the transverse profiles exhibit a pronounced, GRIN-like evolution with strong axial modulation of the central lobe and surrounding rings, consistent with quasi-periodic refocusing/self-imaging. As $\alpha$ increases to $1$ and $10~\mathrm{cm^{-1}}$, the effective focusing becomes progressively concentrated near the entrance face; consequently, the transverse profiles at later planes broaden and the ring contrast decreases, reflecting reduced accumulated GRIN phase and weaker downstream confinement. In the extreme case ($\alpha=50~\mathrm{cm^{-1}}$), the GRIN action is largely confined to a short input region and the profiles beyond $z\gtrsim L/4$ evolve toward near-free-space propagation, leading to comparatively weaker changes with $z$ and a diminished revival of high-contrast ring structure.

\section{Conclusion}
We derived a closed-form ABCD transmission matrix for an end-pumped laser crystal subjected to thermal loading with Beer--Lambert axial absorption. Starting from the steady-state heat equation with a top-hat source term, we obtained the radial temperature distribution and the associated thermo-optic refractive-index profile. Under paraxial conditions, the resulting ray equation reduces to a second-order differential equation with an exponentially varying GRIN strength and admits an analytic solution in terms of Bessel and Neumann functions, leading to compact expressions for the matrix elements. The formulation was validated through comparison with the slab-product method and by recovering the expected limiting cases of a uniform medium and a constant-gradient GRIN medium. In this way, the present treatment replaces repeated axial slicing with a single analytical transfer matrix while retaining the essential physics associated with longitudinal pump absorption.

Beyond the matrix derivation itself, the formalism enables direct propagation analysis of structured beams in a thermally nonuniform crystal. Using the closed-form ABCD elements in the Collins diffraction integral, we obtained an analytical expression for the field of a zero-order Bessel--Gaussian beam and showed that axial absorption has a clear dynamical signature: as the absorption coefficient increases, the thermally induced GRIN action becomes progressively confined to the entrance region of the crystal, the accumulated focusing strength decreases downstream, and the characteristic refocusing and ring contrast of the beam are correspondingly weakened. The results therefore show that longitudinally varying thermal lensing does not merely modify an effective focal length, but can also qualitatively reshape the internal evolution of structured optical fields.

\appendix
\section{Asymptotic behavior}
\label{app:asymptotic}

\subsection{Low-pump-power asymptotics of the closed-form matrix}
\label{subsec:low_pin_limit}

When the pump power is very small, the temperature rise is negligible and the refractive-index
profile approaches a constant, $n(r,z)\simeq n_0$. Since the GRIN strength parameter satisfies
$g\propto S_0\propto P_{\mathrm{in}}$ (see \eqref{eq:grin_strength} and the definition of $S_0$ in \eqref{eq:S0_def}), the limit
$P_{\mathrm{in}}\to 0$ implies
\begin{equation}
g\to 0, \qquad \gamma\equiv \sqrt{g}\to 0, \qquad
p\equiv \frac{2\sqrt{g}}{\alpha}=\frac{2\gamma}{\alpha}\to 0, \qquad
x \equiv p\,u,\quad u\equiv e^{-\alpha z/2}.
\end{equation}
We now take the limit $p\to 0$ directly in the closed-form expressions in Eqs.~\eqref{eq:A_closedform}--\eqref{eq:D_closedform} using the small-argument expansions (Euler's constant $\gamma_E$):
\begin{align}
J_0(t) &= 1-\frac{t^2}{4}+\mathcal{O}(t^4),
&
J_1(t) &= \frac{t}{2}+\mathcal{O}(t^3), \label{eq:smallJ}\\[4pt]
Y_0(t) &= \frac{2}{\pi}\!\left[\gamma_E+\ln\!\Big(\frac{t}{2}\Big)\right]+\mathcal{O}(t^2),
&
Y_1(t) &= -\frac{2}{\pi t}+\mathcal{O}\!\big(t\ln t\big).
\label{eq:smallY}
\end{align}

\paragraph{A-element.}
From \eqref{eq:A_closedform} and using $\gamma=\alpha p/2$,
\begin{equation}
A(z)=\frac{\gamma\pi}{\alpha}\Big[J_{1}(p)\,Y_{0}(x)-Y_{1}(p)\,J_{0}(x)\Big]
=\frac{\pi p}{2}\Big[J_{1}(p)\,Y_{0}(x)-Y_{1}(p)\,J_{0}(x)\Big].
\end{equation}
With Eqs.~\eqref{eq:smallJ}--\eqref{eq:smallY}, the dominant contribution is
$-Y_1(p)J_0(x)\simeq 2/(\pi p)$, hence
\begin{equation}
A(z) \to \frac{\pi p}{2}\cdot \frac{2}{\pi p}=1 \qquad (p\to 0).
\label{eq:A_lowPin}
\end{equation}

\paragraph{B-element.}
From \eqref{eq:B_closedform},
\begin{equation}
B(z)=\frac{\pi}{\alpha}\Big[Y_{0}(p)\,J_{0}(x)-J_{0}(p)\,Y_{0}(x)\Big]
\simeq \frac{\pi}{\alpha}\Big[Y_{0}(p)-Y_{0}(x)\Big],
\end{equation}
since $J_0(p)\simeq J_0(x)\simeq 1$. Using \eqref{eq:smallY},
\begin{equation}
Y_0(p)-Y_0(x)
\simeq \frac{2}{\pi}\ln\!\Big(\frac{p}{x}\Big)
=\frac{2}{\pi}\ln\!\Big(\frac{1}{u}\Big)
=\frac{2}{\pi}\cdot\frac{\alpha z}{2}
=\frac{\alpha z}{\pi},
\end{equation}
where we used $x=p\,u$ and $u=e^{-\alpha z/2}$. Therefore,
\begin{equation}
B(z)\to \frac{\pi}{\alpha}\cdot\frac{\alpha z}{\pi}=z.
\label{eq:B_lowPin}
\end{equation}

\paragraph{C-element.}
From \eqref{eq:C_closedform},
\begin{equation}
C(z)=\frac{\pi g}{\alpha}\,u\Big[J_{1}(p)\,Y_{1}(x)-Y_{1}(p)\,J_{1}(x)\Big].
\end{equation}
Using $J_1(p)\simeq p/2$, $J_1(x)\simeq x/2$, $Y_1(p)\simeq -2/(\pi p)$, $Y_1(x)\simeq -2/(\pi x)$,
one finds the bracketed term is $\mathcal{O}(1)$, so $C(z)=\mathcal{O}(g)\to 0$ as $g\to 0$.
Keeping the leading contribution gives
\begin{equation}
C(z)\simeq \frac{g}{\alpha}\big(e^{-\alpha z}-1\big)
= -\frac{g}{\alpha}\big(1-e^{-\alpha z}\big)
\;\;\xrightarrow[g\to 0]{}\;\;0.
\label{eq:C_lowPin}
\end{equation}

\paragraph{D-element.}
From \eqref{eq:D_closedform} and using $\gamma=\alpha p/2$,
\begin{equation}
D(z)=\frac{\gamma\pi}{\alpha}\,u\Big[Y_{0}(p)\,J_{1}(x)-J_{0}(p)\,Y_{1}(x)\Big]
=\frac{\pi p}{2}\,u\Big[Y_{0}(p)\,J_{1}(x)-J_{0}(p)\,Y_{1}(x)\Big].
\end{equation}
The dominant term is $-J_0(p)Y_1(x)\simeq 2/(\pi x)$, hence
\begin{equation}
D(z)\to \frac{\pi p}{2}\,u\cdot \frac{2}{\pi x}
=\frac{p\,u}{x}=1
\qquad (p\to 0),
\label{eq:D_lowPin}
\end{equation}
because $x=p\,u$.

\subsection{Small-absorption asymptotics of the closed-form matrix}
\label{app:asymptotics}
In this appendix we show directly from Eqs.~\eqref{eq:A_closedform}--\eqref{eq:D_closedform} that, when $\alpha z\ll 1$, the closed-form ABCD elements reduce to the standard constant-gradient GRIN expressions
\begin{equation}
\begin{aligned}
A&\to \cos(\gamma z),\qquad
B\to \frac{1}{\gamma}\sin(\gamma z),\\
C&\to -\gamma\sin(\gamma z),\qquad
D\to \cos(\gamma z),
\end{aligned}
\qquad \gamma=\sqrt{g}.
\end{equation}
Write \eqref{eq:A_closedform} as
\begin{equation}
\begin{aligned}
A(z)&=\frac{\gamma\pi}{\alpha}\Big[J_{1}(p)\,Y_{0}(p\,u)-Y_{1}(p)\,J_{0}(p\,u)\Big],\\
p&=\frac{2\gamma}{\alpha},\qquad u=e^{-\alpha z/2}.
\end{aligned}
\end{equation}
For $\alpha z\ll 1$ we have $u=1-\alpha z/2+\mathcal{O}((\alpha z)^2)$ and therefore
\begin{equation}
p\,u = p-\gamma z+\mathcal{O}(\alpha z^2),\qquad (p\,u-p)=-\gamma z+\mathcal{O}(\alpha z^2).
\end{equation}
Moreover, since $p=2\gamma/\alpha$, fixed $\gamma$ implies $p\gg 1$ as $\alpha\to 0$, and we may use the large-argument asymptotics
\begin{equation}
J_\nu(t)\simeq \sqrt{\frac{2}{\pi t}}\cos\!\Big(t-\frac{\nu\pi}{2}-\frac{\pi}{4}\Big),\qquad
Y_\nu(t)\simeq \sqrt{\frac{2}{\pi t}}\sin\!\Big(t-\frac{\nu\pi}{2}-\frac{\pi}{4}\Big),
\qquad (t\gg 1).
\end{equation}

\paragraph{A-element.}
Substitution yields
\begin{equation}
J_{1}(p)Y_{0}(p u)-Y_{1}(p)J_{0}(p u)\simeq \frac{2}{\pi\sqrt{p(pu)}}\cos(pu-p),
\end{equation}
and therefore
\begin{equation}
\begin{aligned}
A(z)&\simeq \frac{\gamma\pi}{\alpha}\cdot \frac{2}{\pi\sqrt{p(pu)}}\cos(pu-p)
=\frac{2\gamma}{\alpha p\sqrt{u}}\cos(pu-p)\\
&=\frac{1}{\sqrt{u}}\cos(pu-p)
=\cos(\gamma z)+\mathcal{O}(\alpha z).
\end{aligned}
\end{equation}

\paragraph{B-element.}
From \eqref{eq:B_closedform} one finds
\begin{equation}
Y_{0}(p)J_{0}(p u)-J_{0}(p)Y_{0}(p u)\simeq -\frac{2}{\pi\sqrt{p(pu)}}\sin(pu-p),
\end{equation}
hence
\begin{equation}
\begin{aligned}
B(z)&\simeq -\frac{2}{\alpha p\sqrt{u}}\sin(pu-p)
=\frac{1}{\gamma\sqrt{u}}\sin(\gamma z)\\
&=\frac{1}{\gamma}\sin(\gamma z)+\mathcal{O}(\alpha z).
\end{aligned}
\end{equation}

\paragraph{C-element.}
From \eqref{eq:C_closedform},
\begin{equation}
J_{1}(p)Y_{1}(p u)-Y_{1}(p)J_{1}(p u)\simeq \frac{2}{\pi\sqrt{p(pu)}}\sin(pu-p),
\end{equation}
and therefore
\begin{equation}
\begin{aligned}
C(z)&\simeq \frac{\pi\gamma^2}{\alpha}u\cdot \frac{2}{\pi\sqrt{p(pu)}}\sin(pu-p)
=\gamma\sqrt{u}\,\sin(pu-p)\\
&=-\gamma\sin(\gamma z)+\mathcal{O}(\alpha z).
\end{aligned}
\end{equation}

\paragraph{D-element.}
Finally, from \eqref{eq:D_closedform},
\begin{equation}
Y_{0}(p)J_{1}(p u)-J_{0}(p)Y_{1}(p u)\simeq \frac{2}{\pi\sqrt{p(pu)}}\cos(pu-p),
\end{equation}
so
\begin{equation}
\begin{aligned}
D(z)&\simeq \frac{\gamma\pi}{\alpha}u\cdot \frac{2}{\pi\sqrt{p(pu)}}\cos(pu-p)
=\sqrt{u}\cos(pu-p)\\
&=\cos(\gamma z)+\mathcal{O}(\alpha z).
\end{aligned}
\end{equation}
Collecting the results gives the standard constant-gradient GRIN ABCD matrix up to $\mathcal{O}(\alpha z)$ corrections.

\section{Derivation of the Collins diffraction integral in cylindrical coordinates}
\label{app:collins_cyl}
We derive the axially symmetric (cylindrical) form of the Collins diffraction integral starting from its standard Cartesian representation \cite{Pei2019BesselGaussianGRIN}.

For a paraxial optical system characterized by the ray-transfer matrix $\mathbf{M}$ with elements $A(z)$, $B(z)$, $C(z)$, and $D(z)$, the Collins integral in Cartesian coordinates is
\begin{equation}
\begin{aligned}
U_{\mathrm{out}}(x,y,z)
&=\frac{i k}{2\pi B}\iint_{-\infty}^{\infty}
U_{\mathrm{in}}(x_0,y_0)\,\\
&\quad\times \exp\!\left[
-\frac{i k}{2B}
\Big(A(x_0^2+y_0^2)-2(x x_0+y y_0)+D(x^2+y^2)\Big)
\right]
\,\dd x_0\,\dd y_0.
\end{aligned}
\end{equation}
Here, $U_{\mathrm{in}}$ and $U_{\mathrm{out}}$ represent the input and output electric fields, respectively. Introducing polar coordinates $x=r\cos\theta$, $y=r\sin\theta$, $x_0=\rho\cos\phi$, and $y_0=\rho\sin\phi$, with $\dd x_0\,\dd y_0=\rho\,\dd\rho\,\dd\phi$ and $x x_0+y y_0=r\rho\cos(\theta-\phi)$, yields
\begin{equation}
\begin{aligned}
U_{\mathrm{out}}(r,\theta,z)
&=\frac{i k}{2\pi B}
\int_{0}^{\infty}\!\!\int_{0}^{2\pi}
U_{\mathrm{in}}(\rho,\phi)\,\\
&\quad\times \exp\!\left[
-\frac{i k}{2B}
\Big(A\rho^2-2r\rho\cos(\theta-\phi)+Dr^2\Big)
\right]
\,\rho\,\dd\phi\,\dd\rho.
\end{aligned}
\label{eq:collins_cyl_2d}
\end{equation}
For an axially symmetric input $U_{\mathrm{in}}(\rho,\phi)=U_{\mathrm{in}}(\rho)$, the output is independent of $\theta$. The remaining angular integral is
\begin{equation}
\int_{0}^{2\pi}\exp\!\left(i q\cos\phi\right)\,\dd\phi=2\pi J_0(q),
\end{equation}
which reduces the 2D integral to the 1D cylindrical form
\begin{equation}
U_{\mathrm{out}}(r)=\frac{i k}{B}
\exp\!\left(-\frac{i k D}{2B}r^2\right)
\int_{0}^{\infty}
U_{\mathrm{in}}(\rho)\,
\exp\!\left(-\frac{i k A}{2B}\rho^2\right)
J_0\!\left(\frac{k r\rho}{B}\right)\rho\,\dd\rho.
\end{equation}

For completeness, we provide the general integral form of the output field for an $m$th-order BG input.

Starting from the 2D cylindrical Collins integral in \eqref{eq:collins_cyl_2d},
\begin{equation}
\begin{aligned}
U(r,\theta;z)
&=\frac{i k}{2\pi B}\!
\int_{0}^{\infty}\!\!\int_{0}^{2\pi}
U(\rho,\phi;0)\,\\
&\quad\times \exp\!\left[
\frac{i k}{2B}\Big(A\rho^{2}+Dr^{2}-2r\rho\cos(\theta-\phi)\Big)
\right]
\,\rho\, \dd\phi\, \dd\rho,
\end{aligned}
\tag{59}
\end{equation}
and using the input Bessel--Gaussian field of order $m$,
\begin{equation}
U(\rho,\phi;0)=U_0\,J_m(k_r \rho)\,\exp\!\left(-\frac{\rho^2}{w_0^2}\right)\exp(i m\phi),
\end{equation}
The azimuthal integral then takes the form
\begin{equation}
\int_{0}^{2\pi} e^{i m\phi}\exp\!\left(-i\beta\cos(\theta-\phi)\right)\,\dd\phi,
\qquad \beta=\frac{k r\rho}{B}.
\end{equation}
Apply the Jacobi--Anger expansion
\begin{equation}
\exp\!\big(-i \beta \cos\psi\big)=\sum_{\ell=-\infty}^{\infty}(-i)^{\ell}J_{\ell}(\beta)\,e^{i\ell\psi},
\qquad \psi=\theta-\phi,
\end{equation}
and use Fourier orthogonality over $[0,2\pi]$ to obtain
\begin{equation}
\int_{0}^{2\pi} e^{i m\phi}\exp\!\left(-i\beta\cos(\theta-\phi)\right)\,\dd\phi
=
2\pi\,(-i)^m\,e^{i m\theta}\,
J_{m}\!\left(\beta\right).
\end{equation}
Substituting this result into the 2D Collins form reduces it to the single-integral representation
\begin{align}
U(r,\theta;z)
&=\frac{i k}{B}(-i)^m e^{i m\theta}
\exp\!\left(\frac{i k D}{2B}r^2\right)\notag\\
&\quad\times \int_{0}^{\infty}\rho\,J_m(k_r \rho)\,
J_m\!\left(\frac{k r\rho}{B}\right)
\exp\!\left[-\left(\frac{1}{w_0^2}-\frac{i k A}{2B}\right)\rho^2\right]\dd\rho,
\tag{61}
\end{align}
which is the standard Hankel-type form used in the main text.


\section*{Disclosures}
The authors declare no conflicts of interest.

\section*{Data availability}
Data underlying the results presented in this paper are not publicly available at this time but may be obtained from the authors upon reasonable request.

\bibliography{references}

\end{document}